\documentclass[11pt]{article}

 \usepackage{graphicx}
  \usepackage{epsfig}
  \usepackage{amssymb}
  \usepackage{subfigure}

\DeclareMathSymbol{\lesssim}      {\mathrel}{AMSa}{"2E}
\DeclareMathSymbol{\gtrsim}       {\mathrel}{AMSa}{"26}
\begin{document}
\def\be{\begin{equation}}
\def\ee{\end{equation}}
\def\bc{\begin{center}}
\def\ec{\end{center}}
\def\bea{\begin{eqnarray}}
\def\eea{\end{eqnarray}}
\def\dd{\displaystyle}
\def\nn{\nonumber}
\def\ad{\dot{\alpha}}
\def\ov{\overline}
\def\h4w{h_{4 w}}
\def\q4w{q_{4 w}}
\def\hlf{\frac{1}{2}}
\def\qrt{\frac{1}{4}}
\def\as{\alpha_s}
\def\at{\alpha_t}
\def\ab{\alpha_b}
\def\sq2{\sqrt{2}}
\newcommand{\smallz}{{\scriptscriptstyle Z}} %
\newcommand{\mz}{m_\smallz}
\newcommand{\smallw}{{\scriptscriptstyle W}}
\newcommand{\mw}{m_\smallw} 
\newcommand{\smallh}{{\scriptscriptstyle H}}
\newcommand{\mh}{m_\smallh}
\newcommand{\mt}{m_t}
\newcommand{\wh}{w_\smallh}
\newcommand{\wt}{w_t}
\newcommand{\zt}{z_t}
\newcommand{\toh}{t_\smallh}
\def\hto{h_t}
\def\zh{z_\smallh}
\newcommand{\Mvariable}[1]{#1}
\newcommand{\Mfunction}[1]{#1}
\newcommand{\Muserfunction}[1]{#1}
\thispagestyle{empty}
\begin{flushright}
RM3-TH/05-03 \\
CERN-PH-TH/2005-064 
\end{flushright}
\begin{center}
\vspace{1.7cm}
\bc
{\LARGE\bf Two-loop electroweak corrections \\[10pt]
to the Higgs-boson decay  $H \to \gamma \gamma$}

\ec
\vspace{1.4cm}
{\Large \sc Giuseppe Degrassi$^{a}$ and Fabio Maltoni$^{b}$}

\vspace{1.2cm}

${}^a$
{\em 
Dipartimento di Fisica, Universit\`a di Roma Tre\\
INFN, Sezione di Roma III, Via della Vasca Navale~84, I-00146 Rome, Italy}
\vspace{.3cm}

${}^b$
{\em  CERN, CH-1211 Geneva 23, Switzerland}

\end{center}

\vspace{0.8cm}

\centerline{\bf Abstract}
\vspace{2 mm}
\begin{quote}\small
The complete set of two-loop electroweak corrections to the decay width of the
Higgs boson into two photons is presented.  Two-loop contributions
involving  weak bosons and the top quark are computed in
terms of an expansion in the Higgs external momentum.  Adding these
results to the previously known light fermion contributions, we find
that the total electroweak corrections for a Higgs boson with 100 GeV
$\lesssim m_H \lesssim$ 150 GeV are moderate and negative, between $-4
\% \lesssim \delta_{EW} \lesssim 0 \% $.  Combination with the QCD
corrections, which are small and positive, gives a total correction to the
one-loop results of $|\delta_{EW+QCD}| \lesssim 1.5 \%$.  
\end{quote}

\vfill
\newpage
\setcounter{equation}{0}
\setcounter{footnote}{0}
\vskip2truecm
\section{Introduction}
Understanding the mechanism of electroweak symmetry breaking (EWSB) is
one of the main quests of the whole high energy physics community.
The electroweak precision data collected at LEP and SLD in combination
with the direct top-quark mass measurement at the Tevatron, have
strongly constrained the range of possible scenarios and hinted to the
existence of a light scalar particle. Both in the standard model (SM)
and in its minimal supersymmetric extensions (MSSM), the $W$ and $Z$
bosons and fermions acquire masses by coupling to the vacuum
expectation value(s) of scalar SU(2) doublet(s), via the so-called
Higgs mechanism. The striking prediction of such models is the
existence of at least one scalar state, the Higgs boson.  Within the
SM, LEP has put a very strong lower bound to the Higgs mass, $\mh >
114$ GeV~\cite{unknown:2003ih}, and has contributed to build up the
indirect evidence that the Higgs boson should be relatively light with
a high probability for its mass to be below 250 GeV.  In the MSSM the
experimental lower mass bounds for the lightest state are somewhat
weaker but internal consistency of the theory predicts an upper bound
of 140-150 GeV at most~\cite{Allanach:2004rh}.

In this mass intermediate range, 
$80 \lesssim m_H \lesssim 130$ GeV, coupling to
photons even though loop-suppressed and therefore small, is
phenomenologically of great importance. At hadron colliders, the decay
into two photons provides a very clean signature for the discovery in
the gluon-gluon fusion production~\cite{H2gQCD}, for the measurements
of the couplings in the vector-boson fusion
channel~\cite{Rainwater:1997dg} and, depending on the achievable
integrated luminosity, also in the $WH, ZH$, and $t\bar t H$ associated
productions. While none of the above measurements alone can provide
information on the partial width (what is measured is $\sigma(pp \to H)
\cdot$ Br$(H\to \gamma\gamma)$), their combination with signals in other
decay modes, will allow a determination of the total width of the
Higgs and of the couplings with a precision of
10-40\%~\cite{Duhrssen:2004cv}.  A much better determination of the
Higgs width into two photons could be achieved at a $e^+e^-$ linear
collider, via the fusion process $\gamma \gamma \to H$, with the
photons generated by Compton-back scattering of laser
light~\cite{Jikia:1999en,Asner:2001ia}. In this case, 
it has been shown that 
$\sigma(\gamma \gamma \to H) \cdot$ Br$(H\to b \bar b) $ 
could be measured to a precision of a few percents~\cite{Niezurawski:2003iu},
providing an almost direct determination of the width of $H\to \gamma\gamma$  
(the Br$(H\to b \bar b)$ is large for intermediate Higgs 
masses and therefore quite insensitive to the total width).

In view of a precise experimental determination of the 
$H \to \gamma \gamma$ coupling, it is legitimate to ask how well the width can
be predicted in the SM and how sensitive to the effects of new physics
this quantity might be.  The latter question has been the subject of 
several studies~\cite{Kane:1995ek,Djouadi:1996pb,Han:2003gf}.
In general, it is found that corrections to the Higgs width into photons due 
to physics beyond the SM are moderate, ranging up to tens of 
percent.\footnote{This is at variance with the branching ratio into two 
photons which can be drastically modified, due to variations of the total 
Higgs width.} 

The SM prediction for $ \Gamma (H \to \gamma \gamma)$ includes the original 
one-loop  computation~\cite{oneloop}  supplemented by  the
complete two-loop QCD corrections to one-loop top 
contribution~\cite{QCD2loop} and the  two-loop electroweak
corrections evaluated in the large top-mass~\cite{EW2lmt,EW2lkn} and
large Higgs-mass scenarios~\cite{EW2lmh}. 
Recently, also the two-loop contribution  
to $ \Gamma (H \to \gamma \gamma)$  induced by the light fermion
has been computed~\cite{ABDV}.

In this work we present the calculation of the two-loop electroweak 
corrections involving the weak bosons and the top quark
which, together with the previously known contributions 
due to the light fermions~\cite{ABDV}, completes the two-loop determination of 
the $H \to \gamma \gamma$ coupling. Our investigation applies to
the Higgs mass range up to the $2 \,\mw$ threshold covering
the by far most interesting $\mh$ region 
from a phenomenological point of view. 
The paper is organized as follows. In Section 2, we illustrate the
technical details of the calculation focusing on the renormalization procedure
employed. In Section 3 we discuss the numerical results and combine them
with the known two-loop EW light fermion~\cite{ABDV} and two-loop QCD 
corrections~\cite{QCD2loop}. We collect our conclusions in Section 4.

\section{Outline of the calculation}
The general structure of the amplitude for the decay of a Higgs particle
into two photons of polarization vectors 
$\epsilon_\mu (q_1)$ and $\epsilon_\nu (q_2)$,
can be written as:
\be
T^{\mu \nu} = q_1^\mu \, q_1^\nu\, T_1 + q_2^\mu \, q_2^\nu\, T_2 +
              q_1^\mu \, q_2^\nu\, T_3 + q_1^\nu \, q_2^\mu\, T_4 +
              (q_1\cdot q_2) \,g^{\mu \nu} \,T_5 +
         \epsilon^{\mu \nu \rho \sigma} \,q_{1 \rho} \,q_{2 \sigma}\, T_6 \, .
\label{eq:T}
\ee
Abelian gauge invariance requires that $T_1 = T_2 =0$ and $T_4 = -T_5$;
the form factor $T_3$ does not contribute for on-shell photons.
$T_6$ can be generated at the two-loop level, but it has vanishing
interference with the one-loop result. 
%%%%%%%%%%%%%%%%%%%%
\begin{figure}[t]
\begin{center}
\hspace*{0cm}
\epsfig{file=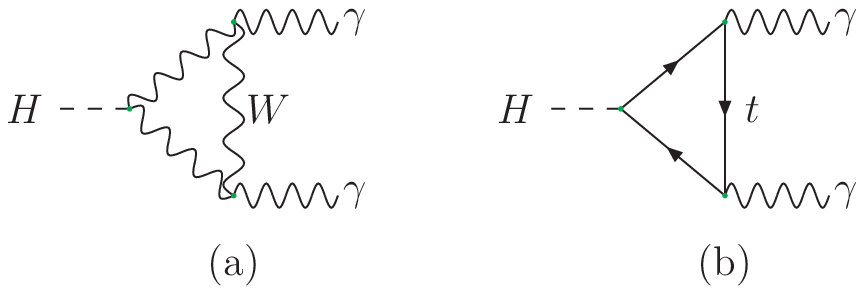,width=10cm}
\vspace*{0cm}
\end{center}
\caption{Representative diagrams of the one-loop contributions to
$H\to \gamma \gamma$.  Diagram (a) represents the bosonic
contributions where the photons couple to charged bosons,  
unphysical scalars and ghosts.  In diagram
(b) the Higgs couples to the fermion line so that only heavy quarks
give a non-negligible contribution.}
\label{fig:one-loop}
\end{figure}
The corresponding partial decay width can be written as:
\bea
\Gamma \left( H \rightarrow \gamma \, \gamma \right) &=&
\frac{G_\mu \alpha^2 \, \mh^3}{128\, \sqrt{2} \, \pi^3}
\left| {\cal F} \right|^2 \, .
\label{eq:G}
\eea Due to the absence of a tree-level Higgs-photon-photon coupling
the lowest order contribution arises at one-loop via $W$ boson and
fermion virtual effects, see Fig.~\ref{fig:one-loop}, the latter almost
entirely due to the top quark with a small correction from the bottom.
The lowest order contribution was computed several years
ago~\cite{oneloop}. Neglecting the bottom part it is given by: 
\bea
{\cal F}^{1l} &=& {\cal F}^{1l}_\smallw + {\cal F}^{1l}_t~,
\label{eq:1loop} \\
{\cal F}^{1l}_\smallw &=& 2\, (1 + 6 \,\wh) - 12\,  \wh \,(1 - 2\, \wh)\,
H \left(-r,-r; -\frac1{\wh} \right)~,  \label{eq:oneloopw} \\
{\cal F}^{1l}_t &=& - 4 Q_t^2 N_c  \,\toh \, \left[ 2 -  
(1 - 4\, \toh )\,
   H \left( -r,-r; -\frac1\toh \right) \right] ~,
\label{eq:onelooptop}
\eea
where $\wh \equiv \mw^2/\mh^2$, $\toh \equiv \mt^2/\mh^2 $,
$N_c$ is the color factor
 and\footnote{All the analytic continuations are obtained with the replacement 
$x \rightarrow x -i\,\epsilon$.}
\be
 H (-r,-r; x ) = \frac12
\log^2 \left( \frac{\sqrt{x+4}-\sqrt{x}}{\sqrt{x+4}+\sqrt{x}}
\right)~.
\label{eq:C0}
\ee
In Eqs.~(\ref{eq:oneloopw}-\ref{eq:C0}) we have expressed the result of the
loop integration in terms of one of the Generalized Harmonic Polylogarithms 
(GHPLs)~\cite{RV} of weight two  employing the notation of Ref.~\cite{AB2}.
At one loop the contribution of light fermions is suppressed by the
smallness of both the Yukawa coupling and the kinematical mass. 
When the Higgs is light, the top
and the bosonic contributions can be expanded 
in $\h4w \equiv  m_H^2/(4\, \mw^2)$ and $h_{4t} \equiv m_H^2/(4\, \mt^2)$ 
with the result
\be
{\cal F}^{1l}_{\smallw} = 
       7 + \frac{22}{15} \h4w 
         + \frac{76}{105}  \h4w^2 
         + \frac{232}{525} \h4w^3  
         + {\cal O}(\h4w^4)\,,
\label{eq:bos-exp}
\ee
and
\be
{\cal F}^{1l}_{t} = - Q_t^2 N_c  
\left(  \frac43 
      + \frac{14}{45} h_{4t} 
      + \frac{8}{63}  h_{4t}^2 
      + \frac{104}{1575} h_{4t}^3  
      + {\cal O}(h_{4t}^4)\right)\,.
\label{eq:top-exp}
\ee
From the above expansions it is manifest that both contributions 
approach constant values (${\cal F}^{1l}_\smallw
\rightarrow 7, \, {\cal F}^{1l}_t \rightarrow  -4/3 Q_t^2 N_c $)
for mass of the particle inside the loop much heavier than $\mh$. 
Furthermore, the $W$ and top one-loop parts are of opposite sign  and
therefore interfere destructively, the former giving the dominant contribution
for light Higgs masses. 
%%%%%%%%%%%%%%%%%%%%%
\begin{figure}[t]
\begin{center}
\vspace*{-.2cm}
\epsfig{file=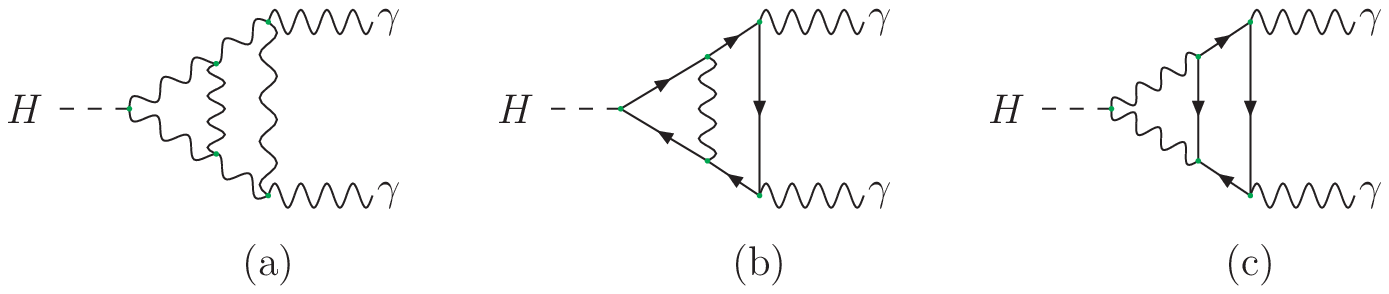,width=15cm}
\end{center}
\caption{The three classes of two-loop diagrams contributing to $H\to
\gamma \gamma$.  Diagram (a) represents purerely bosonic
contributions, which are the corrections to the corresponding diagram
(a) of Fig.~\ref{fig:one-loop}. Diagrams (b) and (c) are both
corrections to the diagram (b) of Fig.~\ref{fig:one-loop}. Leptons and
light quarks start contributing at two loops through diagrams of type
(c).}
\label{fig:two-loop}
\end{figure}
%%%%%%%%%%%%%%%%%%%%%%%%%%%%%%%%%%%%%%%%%

At the two-loop level the electroweak corrections to 
$H  \rightarrow \gamma \gamma $ can be divided in two subsets: 
the corrections induced by the light (assumed massless) fermions and the rest
which involves heavy particles in the loops that can 
be further divided in a purely bosonic contribution and a contribution
involving third generation quarks:
\bea
 {\cal F}^{2l} &=&  {\cal F}^{2l}_{heavy} + {\cal F}^{2l}_{lf}\nn\\
               &=&    {\cal F}^{2l}_\smallw + {\cal F}^{2l}_t
                      + {\cal F}^{2l}_{lf}.
\label{eq:F}
\eea
In Fig.~\ref{fig:two-loop} we  draw one representative diagram for
each type of contribution in $ {\cal F}^{2l}$. We notice that diagrams
of type (c) can also contribute to $ {\cal F}^{2l}_t$ when in the internal
lines a top quark is exchanged. In Eq.~(\ref{eq:F}) 
the last  contribution is very different from the others two because
it involves particles that  do not appear in the one-loop
calculation. Instead, as shown in Fig.~\ref{fig:two-loop}c, 
at the two-loop level, the light fermions may couple 
to the $W$ or $Z$ bosons which in turn  can directly couple to the Higgs 
particle. The light fermion corrections form a gauge invariant subset of
${\cal F}^{2l}$ and have been computed exactly in Ref.~\cite{ABDV},  
where the analytic result has been expressed in 
terms of GHPLs. In that analysis diagrams where 
the bottom quark, which is assumed massless, 
is present together with the $Z$ boson were also included.
%%%%%%%%%%%%%%%%%%%%%
\begin{figure}[t]
\begin{center}
\vspace*{-.1cm}
\epsfig{file=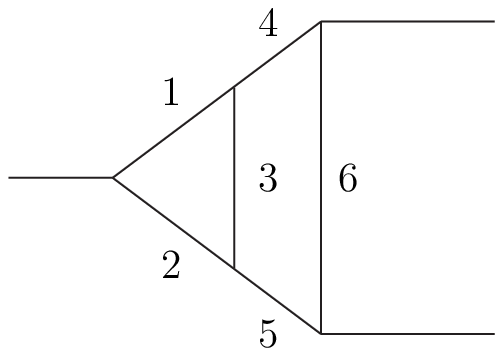,width=5cm}
\vspace*{-.5cm}
\end{center}
\caption{Diagram for studying the cut structure of the amplitude.}
\label{fig:topology}
\end{figure}
%%%%%%%%%%%%%%%%%%%%%%%%%%%%%%%%%%%%%%%%%

As anticipated, in this paper we present the result for ${\cal
F}^{2l}_\smallw$ and ${\cal F}^{2l}_t$ for Higgs mass values in the
intermediate region. Before discussing in detail the structure of the
calculation we notice that for such a values of the Higgs mass the
computation of the one-particle irreducible (1PI) two-loop diagrams
can be obtained via an ordinary Taylor expansion in the variable $\q4w
\equiv q^2/(4\, \mw^2)$ where $q$ is the momentum carried by the
Higgs.  To appreciate this point, we discuss the structure of the cuts
in the Feynman diagrams that contribute to ${\cal F}^{2l}_{heavy}$.
As an example we take the topology drawn in Fig.~\ref{fig:topology},
that is actually present in both sets.  In the ${\cal F}^{2l}_\smallw$
part, the diagrams corresponding to Fig.~\ref{fig:topology} exhibit a
first cut through lines (1,2), (4,5) and (1,3,5) (or (2,3,4)) at $q^2
= 4 \mw^2$ because the only massless particle in the purely bosonic
contribution is the photon (we work in the Feynman gauge) that in this
specific example can only appear in position 3 (see
Fig.~\ref{fig:two-loop}(a) ).  With respect to ${\cal F}^{2l}_t$ it
seems that the diagrams of Fig.~\ref{fig:topology} can develop a cut
at $q^2 = 0$ through lines (4,5) when they represent a bottom quark
(see Fig.~\ref{fig:two-loop}(c)).  However, as discussed in detail in
Ref.~\cite{DM} this cut is actually not present because of the
helicity structure of the diagram. Then, in this set the first cut
arises again at $q^2 = 4 \mw^2$ through lines (1,2).  We notice that
the same topology is actually present also in the light fermion
contribution. In this case because lines 3 and 5 should be taken both
massless the diagrams develop a cut at $q^2 = \mw^2$ through lines
(1,3,5). Indeed the explicit expression for ${\cal F}^{2l}_{lf}$ given
in Ref.~\cite{ABDV} in terms of the GHPLs contains an imaginary part
when $q^2 > \mw^2$.

To evaluate ${\cal F}^{2l}_\smallw $ and  ${\cal F}^{2l}_t$ 
we find it convenient to  employ the Background
Field Method (BFM) quantization framework.
The BFM is a technique for quantizing gauge  theories~\cite{BFM,abbott} that
avoids the complete explicit breaking of the gauge symmetry.
One of the salient features of this approach is that all fields are
split in two components: a classical background field $\hat{V}$
and a quantum field $V$ that appears only in the loops.
The gauge-fixing procedure is achieved through a non linear term in the
fields that breaks the gauge invariance only of the quantum part of the
lagrangian, preserving the gauge symmetry of the effective action with respect
to the background fields. Thus, in the BFM framework
some of the vertices in which  background fields are present are modified
with respect to the standard $R_\xi$ gauge ones. The  complete  set of
BFM Feynman rules for the SM can be found  in Ref.~\cite{ddw}.

In the BFM Feynman gauge (BFG) the heavy two-loop contributions 
to the Higgs decay  into two photons can be organized as
\bea
 {\cal F}^{2l}_{heavy} &=&  K_r {\cal F}^{1l}  +  {\cal F}^{2l}|_{\rm 1PI}\,,
\label{eq:F2}
\eea
where each individual term is finite. In Eq.~(\ref{eq:F2})
the factor $K_r$, whose explicit expression is given in Ref.~\cite{DM},
takes into account the reducible contribution, i.e.,\ the Higgs
wave function renormalization plus the expansion of the bare coupling
$g_0/m_{\smallw_0}$ ($g$ being the $SU(2)$ coupling) in terms of $\mu$-decay
constant, or
\be
K_{r} \equiv  \frac12 \left[  \frac{A_{\smallw \smallw}(0)}{\mw} - V -B +
                      \delta Z_\smallh\, \right]\,,
\label{Kew}
\ee where $A_{\smallw \smallw}(0)$ is the transverse part of the $W$
self-energy at zero momentum transfer, the quantities $V$ and $B$
represent the vertex and box corrections in the $\mu$-decay amplitude
and $\delta Z_\smallh$ is related to the Higgs field wave function
renormalization through \be H_0 = \sqrt{Z_\smallh} H \simeq \left( 1 +
\frac12\,\delta Z_\smallh \right) H~~.  \ee It is known~\cite{ddw,papa} 
that the BFG self-energies coincide with those
obtained in the standard $R_\xi$ gauges via the pinch technique (PT)
procedure~\cite{pinch}. The PT is a prescription that combines the
conventional self-energies with specific parts of the 
vertex and box diagrams, the so-called pinch parts, such that the
resulting PT self-energies are gauge-independent in the class of
$R_\xi$ gauges. Once in Eq.~(\ref{Kew}) the Higgs wave-function 
term $ \delta Z_\smallh$ is intended as the corresponding PT 
quantity~\cite{KPS}, then 
the two terms in Eq.~(\ref{eq:F2}) are actually
finite and gauge-invariant in the $R_\xi$ gauges.
Eq.~(\ref{eq:F2}) can be further divided into a purely bosonic (no
fermionic line present) and a top part 
\bea {\cal F}^{2l}_{\smallw}
&=& K_\smallw {\cal F}^{1l}_\smallw + {\cal F}^{2l}_{\smallw}|_{\rm
1PI}\,,
\label{eq:fw}\\
{\cal F}^{2l}_{t}       &=& ( K_t {\cal F}^{1l} + 
K_\smallw {\cal F}^{1l}_{t}) +  {\cal F}^{2l}_{t}|_{\rm 1PI}\,,
\label{eq:ft}
\eea
where each term is separately finite and gauge
independent. $K_{\smallw,t}$ are the purely bosonic and the top part
of $K_r$ respectively, with\footnote{
In the expression for $K_r$ in Ref.~\cite{DM} $K_t$ corresponds 
to the first line and $K_\smallw$ to the rest.}
$K_\smallw + K_t= K_r$, and ${\cal
F}^{2l}_{\smallw,t}|_{\rm 1PI}$ are the two-loop 1PI corrections plus
the counterterms contribution (apart from the $g_0/m_{\smallw_0}$
factor).

The evaluation of ${\cal F}^{2l}|_{\rm 1PI}$ has been performed via a Taylor
series in $\q4w$ through $O(\q4w^4)$ terms. The  1PI diagrams 
($\sim 1700)$ have been  generated using the program 
FeynArts\footnote{We thanks T. Hahn for 
useful communications.}~\cite{FA}. The relevant form factor, $T_5$, has been
extracted via the use of a standard projector.
The Taylor expansion of the  scalar amplitudes has been obtained
employing an algorithm developed by O.V.~Tarasov~\cite{Tara}. The resulting
two-loop vacuum integrals have been analytically evaluated using the 
expressions of Ref.~\cite{DT}. As a check of our computation we have verified
the  abelian gauge invariance, i.e., $T_1=T_2 =0$. We notice that while
in the standard $R_\xi$ gauges this property is verified only by the on-shell
amplitude, i.e., when $q^2 = \mh^2$, in the BFG it is satisfied also in the
off-shell case, i.e., for arbitrary value of $q^2$.

The tadpole diagrams and the counterterm contribution in  
${\cal F}^{2l}|_{\rm 1PI}$ deserve a  detailed discussion. 
In Eq.~({\ref{eq:G}), the width is expressed in 
terms of $G_\mu$ and $\alpha(0)=1/137.036\dots$, a choice that fixes the 
renormalization of $g$ and of the photon coupling. The other parameters that
require a  renormalization prescription are  the mass of
$W$ boson, of its unphysical  counterpart, $\phi$, and the corresponding
ghost particle, $c$, as well as that of the top quark. Furthermore,
the quartic coupling in the scalar potential, $\lambda$, should also be 
renormalized. In fact $\lambda$ enters in  the $\phi^+ \phi^- h$ coupling 
that is given by $2 \lambda v$, $v$ being the v.e.v.\ of Higgs field.

We employ on-shell  masses for the physical particles.  Then $\delta v$ is
fixed in terms of $\delta g$ and $\delta \mw^2$. In the Feynman gauge we 
use, the mass renormalization for the $c$ and $\phi$ fields  can be chosen to 
be equal to that of the $W$ mass, i.e., 
$\delta m^2_c =  \delta \mw^2,\: \delta m^2_\phi = \delta \mw^2$.

We eliminate  the tadpole diagrams by fixing the tadpole counterterm  to 
cancel them. This  implies that the renormalization of the $\phi$ mass   
should be augmented by the tadpole contribution, $\delta \tau$, i.e., 
\be
\delta m^2_\phi = \delta \mw^2 + \delta \tau = 
{\rm Re}\, A_{\smallw \smallw}(\mw^2) -\frac{T}v\,,
\label{eq:phimass}
\ee  
where  $i\, T$ is the sum of 1PI tadpole diagrams with external leg extracted. 

The renormalization of $\lambda$  is achieved following the prescription 
given in  Ref.~\cite{SZ} for the renormalization of the Higgs sector. 
However, once the factor $g/m_{\smallw}$ is extracted and its renormalization
included in the $K_r$ term, the relevant coupling becomes 
$ ( m_{\smallw}/g)\, 2 \lambda v = \lambda v^2 $ whose counterterm is equal to
\be
\delta ( \lambda v^2) = \frac12 \left( \delta \mh^2 - \delta \tau \right)
= \frac12 \left( {\rm Re} \,\Pi_{HH}(\mh^2) + \frac{T}v \right)\,,
\ee
where $\Pi_{HH}$  is the  Higgs self-energy.

The structure of the counterterms discussed above is sufficient to
obtain finite results at the S-matrix level, namely when the amplitude
is evaluated on shell, i.e., $q^2 = \mh^2$. We are, instead,
evaluating ${\cal F}^{2l}|_{\rm 1PI} $ via a Taylor series in $\q4w$
and therefore actually computing an off-shell amplitude that only at
the end will be taken on the mass-shell, i.e., only at the end we are
going to let $\q4w \to \h4w$. The set of counterterms specified above
is not sufficient to make each of the individual term in the $\q4w$
expansion finite although the total sum is (to the order of the
expansion) as it should be. Indeed it is known that, beyond one-loop,
in order to obtain finite background-field vertex functions the
renormalization of the quantum gauge parameter, $\xi_Q$ is also
needed~\cite{abbott}.  Clearly, for S-matrix elements, which are
independent upon the gauge parameter, this renormalization is
irrelevant. It should be said that the renormalization of the gauge
parameter can be avoided if one employs the Landau gauge, $\xi_Q
=0$. However, this gauge is not very practical in the BFM because of
the presence in the three and four gauge-boson vertices of terms
proportional to $1/\xi_Q$~\cite{ddw}. Then in this gauge an arbitrary
gauge parameter should be retained until all the $1/\xi_Q$ terms have
been canceled.
%%%%%%%%%%%%%%%%%%%%%
\begin{figure}[t]
\begin{center}
\vspace*{0cm}
\epsfig{file=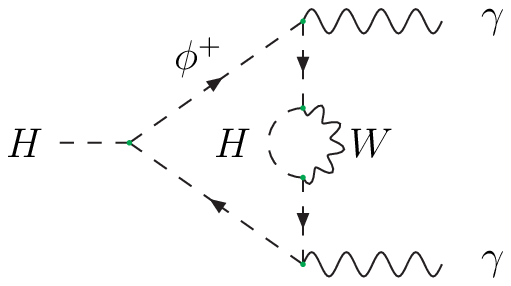,width=5cm}
\vspace*{0cm}
\end{center}
\caption{A two-loop diagram containing a 
quadratically divergent subdiagram (self-energy) 
associated to the unphysical scalars $\phi$.}
\label{fig:self}
\end{figure}
%%%%%%%%%%%%%%%%%%%%%%%%%%%%%%%%%%%%%%%%%
The renormalization of the gauge parameter is needed to obtain the
finiteness of each of the individual terms in the $\q4w$ expansion. In
fact, diagrams where the quantum $\phi$ field is exchanged can contain
quadratically divergent subdiagrams associated to the $\phi$
self-energy where the Higgs field is present, 
see Fig.~\ref{fig:self}. To make these diagrams finite
one needs a further subtraction proportional to the derivative of the
quantum $\phi$ field self-energy. The counterterm for the gauge
parameter that makes each individual term of the Taylor expansion
finite is 
\be
\delta \xi_Q = -{\delta \mw^2 \over \mw^2}  + \delta Z_\phi\,,
\label{eq:xicont}
\ee
where $ \delta Z_\phi$ is  the derivative of the   $\phi$ self-energy 
evaluated at zero momentum transfer. Few observations are now in order.
i) As always only the divergent part of $\delta \xi_Q$ is fixed. Our choice in
Eq.~(\ref{eq:xicont}) specifies the finite term. As said before,
S-matrix elements are insensitive to the renormalization of the 
gauge parameter and 
therefore to the prescription used for it. However, in our actual calculation
the expansion in $\q4w$ includes some higher order terms in $\h4w$
and therefore our result contains  a residual dependence on the 
prescription for $\delta \xi_Q$. ii) We notice that the first term in the
r.h.s.~of Eq.~(\ref{eq:xicont}) has the effect to cancel the Feynman gauge
mass renormalization for the $c$ and 
$\phi$ we have previously introduced, or
 \be
\left. \delta m^2_c = \delta (\xi_Q \mw^2) \right|_{\xi_Q =1} = 
\left. \mw^2 \,\delta \xi_Q + \xi_Q \,\delta \mw^2 \right|_{\xi_Q =1} =  
\delta Z_\phi \mw^2~.
\ee 
Similarly,  the $W$ boson propagator  is 
renormalized,  a part longitudinal terms proportional to $ \delta Z_\phi$,
as if we were employing in the one-loop part the  Landau gauge expression for 
it.

\section{Numerical Results}
In this section we present the result of our computation. 
As explained in the previous section the evaluation of  the 1PI contributions 
${\cal F}^{2l}_{t,\smallw} |_{1PI}$ has been obtained by expanding the 
two-loop diagrams in terms of the variable $\q4w$, or
\be
{\cal F}^{2l}_{t,\smallw} |_{1PI} =  \frac{\alpha}{4 \pi s^2}
\left( c_0 + c_1 \q4w + c_2 \q4w^2 + c_3 \q4w^3 + {\cal O}(\q4w^4) \right) \,.
\label{eq:exp}
\ee
The coefficients $c_i,\, (i=0,..,3)$   depend on  $\mh$.  Their 
analytic expressions are too long to be reported here, therefore 
we present them in a numerical form. 
The $c_i$ coefficients  for 
100 GeV $< \mh<$ 150 GeV are very well described by a linear fit in $\h4w$.  
Choosing  $\mt= 178$ GeV, $\mw = 80.4$ GeV and $\mz= 91.18$ GeV, 
$s^2=1-\mw^2/\mz^2$,  we obtain for the heavy-quark contribution, 
${\cal F}^{2l}_{t}|_{1PI}$,
\bea
c_0 &=&-54.4 + 6.07 \, \h4w \nn \\
c_1 &=&-13.3 + 3.02 \, \h4w \nn \\
c_2 &=&-7.00 + 1.84 \, \h4w \nn\\
c_3 &=&-4.35 + 1.18 \, \h4w \nn~, 
\eea
while  for the purely bosonic one, ${\cal F}^{2l}_{\smallw}|_{1PI}$, we have
\bea
c_0 &=&16.3 - 1.72 \, \h4w \nn \\
c_1 &=&25.7 - 2.64 \, \h4w \nn \\
c_2 &=&15.5 - 2.05 \, \h4w \nn \\
c_3 &=&10.2 - 1.46 \, \h4w \nn~.
\eea
%%%%%%%%%%%%%%%%%%%%%%%%%%%%%%%%%%%%%%%%%%%%%%%%%%%%%%%%%%%%%%%%%%%%%%%%%%%%%
\begin{table}[t]
\addtolength{\arraycolsep}{0.1cm}
\renewcommand{\arraystretch}{1.4}
\begin{center}
\begin{tabular}[4]{c|cc|cc}
\hline
\hline
$m_h$ & \multicolumn{2}{c|}{${\cal F}^{2l}_{\smallw}|_{1PI}$} & 
                            \multicolumn{2}{c}{${\cal F}^{2l}_{t}|_{1PI}$} \\
  & truncated & Pad\'e & truncated & Pad\'e \\
\hline
100 & 27.8 & 27.9 & -57.9 & -57.9 \\
105 & 29.3 & 29.6 & -58.3 & -58.4 \\
110 & 31.1 & 31.5 & -58.8 & -59.0 \\
115 & 32.9 & 33.6 & -59.3 & -59.6 \\
120 & 35.0 & 35.9 & -59.9 & -60.2 \\
125 & 37.2 & 38.5 & -60.5 & -61.0 \\
130 & 39.6 & 41.5 & -61.2 & -61.9 \\
135 & 42.2 & 45.0 & -62.0 & -62.9 \\
140 & 45.1 & 48.9 & -62.8 & -64.1 \\
145 & 48.1 & 53.6 & -63.7 & -65.5 \\
150 & 51.4 & 59.2 & -64.7 & -67.2 \\
\hline
\hline
\end{tabular}
\end{center}
\caption{Comparison between normal truncated Taylor expansions of 
${\cal F}^{2l}_{t,\smallw}|_{1PI}$ and Pad\'e improved values obtained using
Eq.~(\ref{eq:pade}). Numbers are given in units of $\alpha/(4 \pi s^2)$.}
\label{tab:one}
\end{table}
%%%%%%%%%%%%%%%%%%%%%%%%%%%%%%%%%%%%%%%%%%%%%%%%%%%%%%%%%%%%%%%%%%%%%%%%%%%%%%%
From the above expressions it is clear that the heavy-quark
contribution shows a very good convergence for Higgs masses in the
intermediate region while the purely bosonic expansion has a slightly
worse behaviour.  In both cases, however, the series behave better
than a geometric series.  In order to improve the convergence of our
expansion close to the $2 m_\smallw$ threshold, i.e., to estimate the
impact of the higher order terms, we employ a Pad\'e approximant.
This method has been shown to be a very powerful tool to obtain an
approximation to an analytic function $f(x)$ which cannot be computed
directly, but it is known for small (and/or large) argument.\footnote{
For a short review see Ref.~\cite{Harlander:2001sa}.}  The
generic Pad\'e approximant, $P_{[n/m]}(x)$ is the ratio between two
polynomials of degree $n$ and $m$, respectively.  It is known that
best convergence is achieved when the polynomial in the numerator has
degree equal to or greater by one than the polynomial in the
denominator, so we choose\footnote{In order to gain confidence in
our method we checked our procedure on the one-loop expansions,
Eqs.~(\ref{eq:bos-exp},\ref{eq:top-exp}), and compared with the exact
results, Eqs.~(\ref{eq:oneloopw},\ref{eq:onelooptop}). We also checked
that other choices for the degrees of the polynomials give similar
results.}  \be P_{[2/1]}(x)= \frac{a_0 + a_1 x + a_2 x^2}{1+ b_1 x}
\label{eq:pade}\,.  \ee The coefficients $a_i$ and $b_i$ are found by
matching the Taylor expansion of Eq.~(\ref{eq:pade}) to the
coefficients $c_i$ of the expansion of Eq.~(\ref{eq:exp}).  This
procedure yields the following set of equations: 
\bea 
&&c_0 = + a_0 \nn \\ 
&&c_1 = - a_0 b_1   + a_1  \nn \\ 
&&c_2 = + a_0 b_1^2 - a_1 b_1   + a_2 \nn\\ 
&&c_3 = - a_0 b_1^3 + a_1 b_1^2 - a_2 b_1 \nn\,,
\label{eq:matching}
\eea
which can be easily solved in terms of $a_{0,1,2}$ and $b_1$.
Numerical results are shown in Tab.~\ref{tab:one}, where the fixed order
Taylor expansion and the Pad\'e approximants are given for different 
Higgs masses. As expected the impact of the improvement is larger close to the 
$2 m_\smallw$ threshold entailing a $15\%$ enhancement for the bosonic 
expansion and only a $4\%$ for the heavy-quark one. Note that 
the size of these effects is consistent with estimating the error 
of the results by using the last coefficient in the Taylor expansion. 

%%%%%%%%%%%%%%%%%%%%%%%%%%%%%%%%%%%%%%%%%%%%%%%%%%%%%%%%%%%%%%%%%%%%%%%%%%%%%
\begin{table}[t]
\addtolength{\arraycolsep}{0.1cm}
\renewcommand{\arraystretch}{1.4}
\begin{center}
\begin{tabular}[4]{c|ccccc|c}
\hline
\hline
$m_h$ & leptons & lq & $3^{\rm rd}$ gen & YM & 2-loop & 
                                                         $\delta_{EW} (\%)$ \\
\hline
100   & -8.04 & -10.9 & -30.9 & 13.5 & -36.3 & -3.44\\ 
105   & -8.07 & -10.6 & -31.3 & 15.5 & -34.5 & -3.22\\ 
110   & -8.07 & -10.3 & -31.7 & 17.6 & -32.4 & -2.97\\ 
115   & -8.04 & -9.86 & -32.1 & 20.0 & -30.0 & -2.70\\ 
120   & -7.95 & -9.29 & -32.6 & 22.5 & -27.2 & -2.40\\ 
125   & -7.82 & -8.59 & -33.1 & 25.4 & -24.1 & -2.07\\ 
130   & -7.62 & -7.75 & -33.6 & 28.5 & -20.4 & -1.71\\ 
135   & -7.35 & -6.73 & -34.2 & 32.0 & -16.2 & -1.32\\ 
140   & -6.98 & -5.47 & -34.7 & 35.8 & -11.3 & -0.88\\ 
145   & -6.48 & -3.89 & -35.2 & 40.0 & -5.61 & -0.42\\ 
150   & -5.78 & -1.78 & -35.5 & 44.1 & 1.022 & 0.072\\ 
\hline
\hline
\end{tabular}
\end{center}
\caption{Contributions to ${\cal F}$ 
at two loops, in units of $\alpha/(4 \pi s^2)$, for various Higgs masses.
Starting from the second column, the two-loop contributions 
of the following classes of diagrams are shown, 
as listed in Eq.~(\ref{eq:2-loop-contributions}): 
sum of three lepton families, 
light quarks ($u,d,c,s$), 
third generation quarks, 
purely bosonic. The sixth column shows the sum of all the 
2-loop EW contribution, ${\cal F}^{2l}$. The last column gives 
the total EW correction to the decay rate as plotted in Fig.~\ref{fig:delta}.}
\label{tab:two}
\end{table}
%%%%%%%%%%%%%%%%%%%%%%%%%%%%%%%%%%%%%%%%%%%%%%%%%%%%%%%%%%%%%%%%%%%%%%%%%%%%%%%
%%%%%%%%%%%%%%%%%%%%%%%%%%%%%%%%%%%%%%%%%%%%%%%%%%%%%%%%%%%%%%%%%%%%%%%%%%%%%%%
\begin{figure}[t!]
\begin{center}
\vspace*{-1.7cm}
\hspace*{-1cm}\epsfig{file=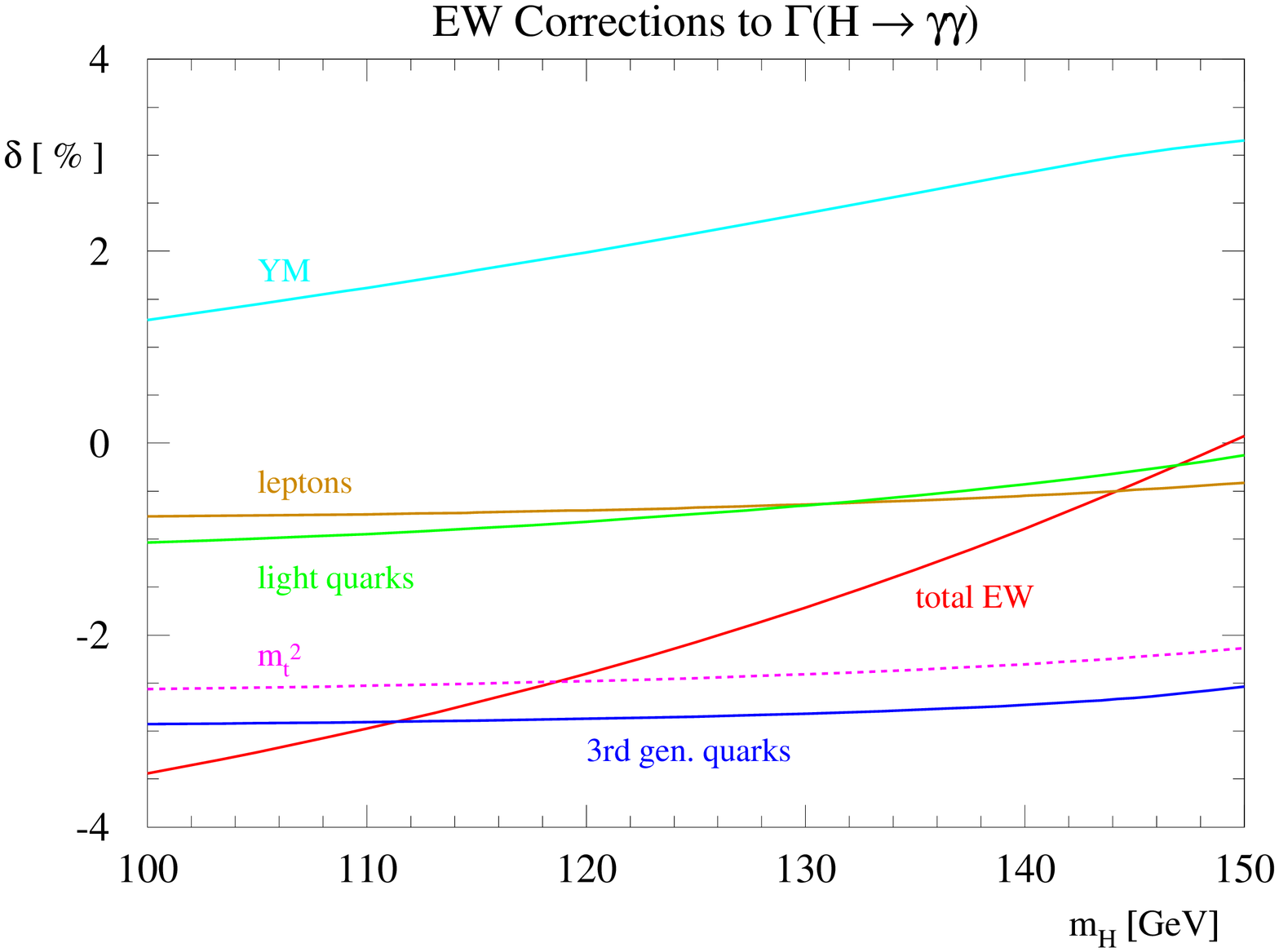,angle=-90,width=18cm}
\vspace*{-1.2cm}
\end{center}
\caption{Various contributions to $\delta_{EW}$ as a function of the
Higgs mass.  Lepton (summed over three families) and light quark
contributions ($u,d,c,s$) are the two central curves. Purely bosonic
(YM) and third generation quarks are the top and the bottom curves
respectively.  The large top-mass approximation ($\mt^2$), which is a
subset of the third generation contribution, is also shown (dotted
line). }
\label{fig:EWcontributions}
\end{figure}
%%%%%%%%%%%%%%%%%%%%%%%%%%%%%%%%%%%%%%%%%%%%%%%%%%%%%%%%%%%%%%%%%%%%%%%%%%%%%%%

Our result on the heavy corrections can be put together with the
result of Ref.~\cite{ABDV} on the light fermion contribution to obtain
a complete prediction for the two-loop electroweak correction to the
decay width, \vskip0.5cm
\begin{tabular}{cll}
${\cal F}^{2l} = \frac{\alpha}{4 \pi s^2}\{$&
$+ 3 (\wh C_W^{l} A_2[0   ,\wh]+ z_H C_Z^{l} A_1[\zh])$ & leptons\\[10pt]
&$+ 2 (\wh C_W^{q} A_2[-2/9,\wh]+ z_H C_Z^{q} A_1[\zh])$ & light quarks\\[10pt] 
&$+ \zh C_Z^{b}  A_1[\zh] + {\cal F}^{2l}_t $ & third generation quarks\\[10pt]
&$+ {\cal F}^{2l}_\smallw  \left\} \right. $ & YM\,,
\end{tabular}
\vskip-0.5cm
\be
~
\label{eq:2-loop-contributions}
\ee
where $A_1[x]$ and $A_2[q,x]$ are defined in Ref.~\cite{ABDV}, 
$\zh \equiv \mz^2/\mh^2$ and 
\bea
C_Z^{q} &=&  \frac{4N_c}{c^2} 
\left[ Q_u^2 ( {z_-^u}^2 + {z_+^u}^2 ) 
     + Q_d^2 ( {z_-^d}^2 + {z_+^d}^2 )\right] \nn \\
C_Z^{l} &=&  \frac{4}{c^2} ( {z_-^l}^2 + {z_+^l}^2 )\nn \\
C_Z^{b} &=&  \frac{4N_c}{c^2} Q_d^2 ( {z_-^d}^2 + {z_+^d}^2 ) \\
C_W^{q} &=& 2 N_c\nn \\
C_W^{l} &=& 2 \nn\,,
\eea
with  $z_+^i = T_3 - Q_i s^2$ and $z_-^i = - Q_i s^2$.
%%%%%%%%%%%%%%%%%%%%%%%%%%%%%%%%%%%%%%%%%%%%%%%%%%%%%%%%%%%%%%%%%%%%%%%%%%%%%%%
\begin{figure}[t!]
\begin{center}
\vspace*{-1.7cm}
\hspace*{-1cm} \epsfig{file=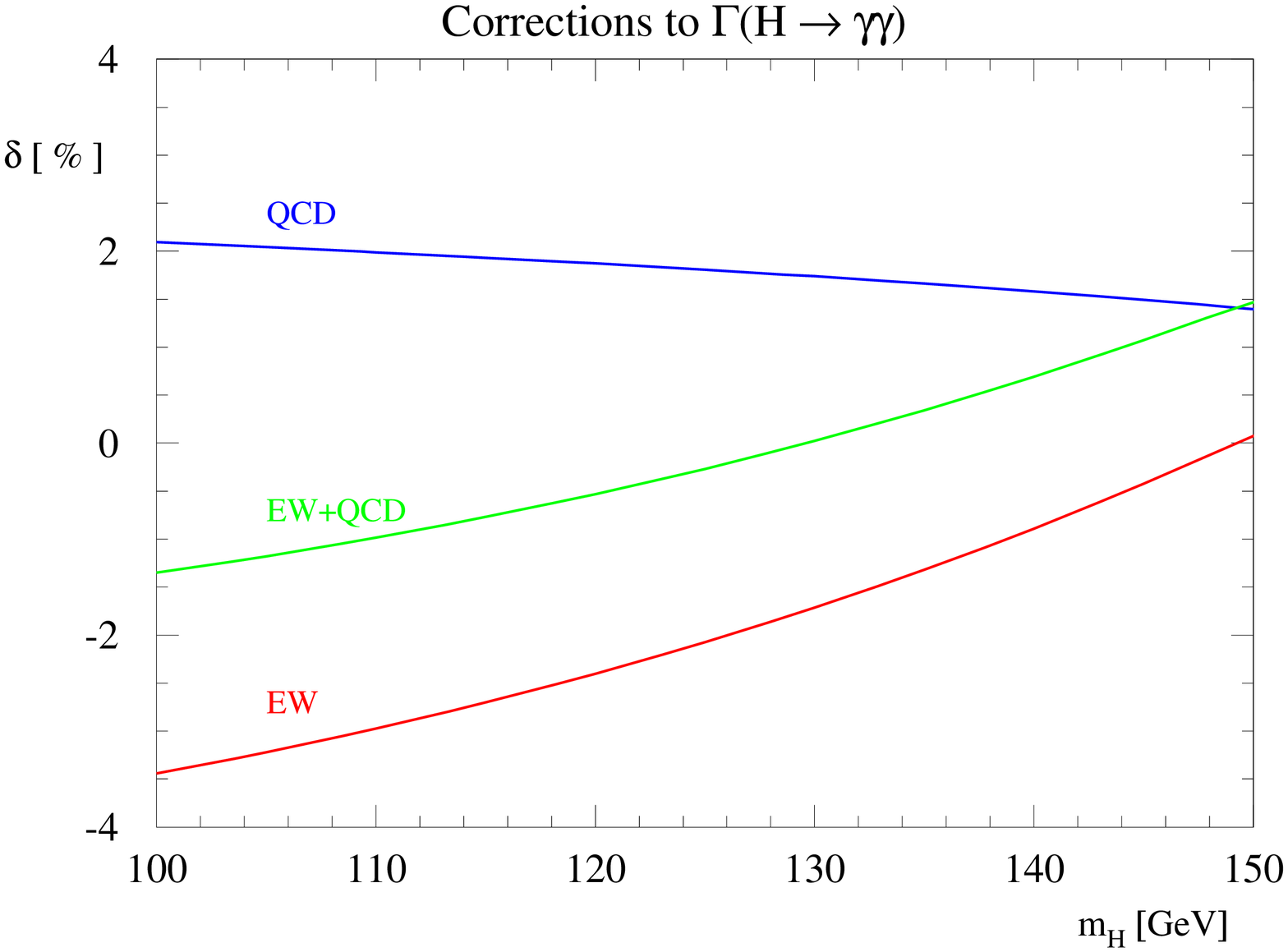,angle=-90,width=18cm}
\vspace*{-1.2cm}
\end{center}
\caption{Corrections to the decay rate of $\Gamma(H \to \gamma \gamma)$ for 
various Higgs masses. 
The upper curve corresponds to the QCD corrections, 
the lower curve represents the complete electroweak corrections.
Their sum is given by the intermediate curve. }
\label{fig:delta}
\end{figure}
%%%%%%%%%%%%%%%%%%%%%%%%%%%%%%%%%%%%%%%%%%%%%%%%%%%%%%%%%%%%%%%%%%%%%%%%%%%%%%%
The numerical impact of each of the contributions in
Eq.~(\ref{eq:2-loop-contributions}) is shown in Tab.~\ref{tab:two} .
As a generic feature we note that the two-loop contributions involving
fermions in the loop are negative, while purely bosonic contributions
are positive, in analogy to the one-loop calculation. For Higgs masses
around 100 GeV the dominant effect comes from the third generation
quarks, with the purely bosonic canceling most of the lepton and
light quarks contributions. For higher values of the Higgs mass, the
purely bosonic term increases and becomes comparable to the top
contribution but with opposite sign. In this region a strong
cancellation between the two leading terms takes place, leaving a very
small correction as a final result.  The corresponding corrections to
the width $\Gamma (H \to \gamma \gamma ) = \Gamma_0 \cdot ( 1 +
\delta_{EW})$ can be calculated as 
\be 
\delta_{EW} = \frac{2 {\rm
Re}({\cal F}^{1l} {\cal F}^{2l})}{|{\cal F}^{1l}|^2} \,, 
\ee 
and are
shown in Fig.~\ref{fig:EWcontributions}.

From our expansions it is easy to extract the leading term in $G_\mu m_t^2$,
which was calculated in Refs.~\cite{EW2lkn}. We find
\be
\lim_{m_t \to \infty}  
{\cal F}^{2l}_t = -\frac{\alpha}{4 \pi s^2} N_c Q_t^2 \frac{\mt^2}{\mw^2} 
\left( 
\frac{367}{96}  
+\frac{11}{16} \h4w 
+\frac{19}{56} \h4w^2 
+\frac{29}{140} \h4w^3 + {\cal O}(\h4w^4) \right)\,.
\ee
The contribution from this (gauge invariant) class of electroweak 
corrections is also shown in Fig.~\ref{fig:EWcontributions}. 
The first important observation is that indeed 
the leading term in $G_\mu m_t^2$ approximates quite well the contribution
from the third generation quarks in the whole range of Higgs masses between
100 GeV and 150 GeV. However, as shown in Fig.~\ref{fig:EWcontributions}, 
this contribution is never the dominant one. The fact that it approximately 
reproduces the total electroweak corrections for Higgs masses around 120 GeV 
is due to a fortuitous cancellation between the purely bosonic and the 
light quark and lepton terms. In fact,  for Higgs masses above 140 GeV, 
the $G_\mu m_t^2$ contribution is mostly canceled by the purely bosonic one 
and therefore it is much larger than the total electroweak correction. 

Finally, it is interesting to compare and combine the 
total electroweak correction with the QCD one.
As a check of our techniques we have recomputed it 
as an expansion in terms of $h_{4t}$, obtaining
\be
{\cal F}^{2l}_{QCD} = \frac{\alpha_S}{\pi} \frac{4 Q_t^2 N_c}{3}   
\left( 1 - \frac{122}{135} h_{4t} 
         - \frac{8864}{14175}  h_{4t}^2 
         - \frac{209186}{496125} h_{4t}^3  + {\cal O}(h_{4t}^4)\right)\,,
\ee
in complete agreement with the known results~\cite{QCD2loop}. We use the above
expansion in our numerical analysis, since it converges very rapidly
for Higgs masses in the range we are interested in. The impact of the
QCD corrections, shown in Fig.~\ref{fig:delta}, is small and amounts
to an increase of about 2\% of the decay width. Such a small
contribution is expected since for intermediate Higgs masses, the 
one-loop result is dominated by the bosonic loop, which is unaffected by  
QCD effects.  Due the difference in sign between the EW and QCD
contributions, the total correction $\delta_{EW+QCD}$ turns out to be very
small, ranging between $-1.5 \%$ for $m_H=100$ GeV to $1.5 \%$ for $m_H=150$ 
GeV and reaching an almost perfect cancellation around $\mh=130$ GeV.

\section{Conclusions}
We have computed the two-loop electroweak corrections to the decay
width of the Higgs into two photons induced by the weak bosons and
third generation quarks. By combining our results with those of
Ref.~\cite{ABDV}, involving light fermions in the loops, we have found
that the total electroweak corrections to the decay rate are moderate
and negative. For Higgs masses between 100 GeV $\lesssim \mh \lesssim$
150 GeV they range $-4 \% \lesssim \delta_{EW} \lesssim 0 \% $. Once
QCD corrections, which are also small but positive, are added, one
finds that $|\delta_{EW+QCD}|$ is always less than $1.5 \%$.  This
shows that the perturbative expansions in $\alpha_S$ and $\alpha_{EW}$
for the decay rate are extremely well behaved and a next-to-leading
calculation already gives a very reliable prediction.  If a similar
precision could be matched experimentally, one would have an
interesting and powerful test of the standard model.

\section{Acknowledgments}
We are grateful to Alessandro Vicini for useful discussions.

\end{document}